\documentclass[aps,prl,twocolumn,groupedaddress]{revtex4}
\usepackage{amsmath}
\usepackage{graphicx}

\newcommand\MET{\mbox{$\slash\hspace{-0.65em}E_T$}}

\begin{document}



\title{
Measurement of the {\boldmath$WW$} production cross section 
with dilepton final states in {\boldmath$p\bar{p}$} collisions at {\boldmath$\sqrt{s}=1.96$}~TeV
and limits on anomalous trilinear gauge couplings
}

%
\author{V.M.~Abazov$^{37}$}
\author{B.~Abbott$^{75}$}
\author{M.~Abolins$^{65}$}
\author{B.S.~Acharya$^{30}$}
\author{M.~Adams$^{51}$}
\author{T.~Adams$^{49}$}
\author{E.~Aguilo$^{6}$}
\author{M.~Ahsan$^{59}$}
\author{G.D.~Alexeev$^{37}$}
\author{G.~Alkhazov$^{41}$}
% ...

\affiliation{\vspace{0.1 in}(The D\O\ Collaboration)\vspace{0.1 in}}
% ...
\affiliation{$^{6}$University of Alberta, Edmonton, Alberta, Canada;
                Simon Fraser University, Burnaby, British Columbia, Canada;
                York University, Toronto, Ontario, Canada and
                McGill University, Montreal, Quebec, Canada}
% ...
\affiliation{$^{30}$Tata Institute of Fundamental Research, Mumbai, India}
% ...
\affiliation{$^{37}$Joint Institute for Nuclear Research, Dubna, Russia}
% ...
\affiliation{$^{41}$Petersburg Nuclear Physics Institute,
                St. Petersburg, Russia}
% ...
\affiliation{$^{49}$Florida State University, Tallahassee, Florida 32306, USA}
% ...
\affiliation{$^{51}$University of Illinois at Chicago,
                Chicago, Illinois 60607, USA}
% ...
\affiliation{$^{59}$Kansas State University, Manhattan, Kansas 66506, USA}
% ...
\affiliation{$^{65}$Michigan State University,
                East Lansing, Michigan 48824, USA}
% ...
\affiliation{$^{75}$University of Oklahoma, Norman, Oklahoma 73019, USA}
% ...

\collaboration{D0 Collaboration}

\date{April 3, 2009}

\begin{abstract}
We provide the most precise measurement of the $WW$ production cross section in 
$p \bar{p}$ collisions to date at a center of mass energy of 1.96~TeV, 
and set limits on the associated trilinear gauge couplings.
The \mbox{$WW \to \ell\nu\ell^\prime\nu$} ($\ell,\ell^\prime=e,\mu$) decay channels are analyzed
in 1~fb$^{-1}$ of data collected by the D0 detector at the Fermilab Tevatron Collider.
The measured cross section is
$\sigma(p\bar{p}\to WW) = 11.5\ \pm  2.1\text{ (stat + syst)}\ \pm 0.7 \text{ (lumi) pb}$.
One- and two-dimensional 95\%~C.L.\ limits on trilinear gauge couplings are provided.
\end{abstract}


\maketitle

The non-Abelian gauge group structure of the electro\-weak sector of the
standard model (SM) predicts specific interactions between the $\gamma$,
$W$, and $Z$ bosons.  Two vertices, $WW\gamma$ and $WWZ$, 
provide important contributions to the \mbox{$p\bar{p}\to WW$}  production cross section. 
Understanding this process is imperative because it is an irreducible background 
to the most sensitive discovery channel for the Higgs boson at the Tevatron, $H\to WW$. 
A detailed study of $WW$ production also
probes the triple gauge-boson couplings (TGCs), which are sensitive to low-energy
manifestations of new physics from a higher mass scale,
and is sensitive to the production and decay of new particles,
such as the Higgs boson~\cite{Hagiwara}. 
Studying $WW$ production at the Fermilab Tevatron Collider provides
an opportunity to explore constituent center of mass energies ($\sqrt{\hat{s}}$) higher than
that available at the CERN $e^+e^-$ Collider (LEP)~\cite{LEP}, since SM $WW$ production 
at the Tevatron has an average $\sqrt{\hat{s}} = 245$ GeV and a 57\% 
probability for $\sqrt{\hat{s}} > 208$ GeV~\cite{Hagiwara}.
The Tevatron experiments have been active in studying the $WW$ cross section and TGCs
in the past~\cite{d0RunII,cdfww,d0wwac}.
In this Letter we present the most precise measurement of the $WW$ production cross section 
in $p\bar{p}$ collisions to date and updated limits on anomalous $WW\gamma$ and $WWZ$ couplings.

We examine $WW$ production
via the process $p\bar{p}\to W^+W^-\to\ell^+\nu\ell^{\prime-}\bar{\nu}$
($\ell,\ell^\prime=e,\mu$; allowing for $W\to\tau\nu\to\ell+n\nu$ decays)
and use charged lepton transverse momentum ($p_T$) distributions to 
study the TGCs.
The decay of two $W$ bosons into electrons or muons results in
a pair of isolated, high-$p_T$, oppositely charged leptons 
and a large amount of missing transverse energy (\MET) due to the escaping neutrinos.  
This analysis uses $p \bar{p}$ collisions at a center of mass energy of 1.96~TeV, as 
recorded by the D0 detector~\cite{nim} at the Tevatron.
A combination of single-electron ($ee$ and $e\mu$ channels) or 
single-muon ($\mu\mu$ channel) triggers were used to collect
the data, which correspond to integrated luminosities 
of 1104 ($ee$), 1072 ($e\mu$), and 1002 ($\mu\mu$)~pb$^{-1}$~\cite{lumierr}.  

Electrons are identified in the calorimeter by their electromagnetic showers,
which must occur within $|\eta|<1.1$ or $1.5<|\eta|<3.0$~\cite{d0coord}.  
In the $ee$ channel, at least one electron must satisfy $|\eta|<1.1$.
Electron candidates must be spatially matched to a track from the central 
tracking system, isolated from other energetic particles,  
and have a shape consistent with that of an electromagnetic shower.  
Electron candidates must also satisfy
a tight requirement on a multivariate electron discriminant which takes into
account track quality, shower shape, calorimeter and track isolation,
and $E/p$, where $E$ is the calorimeter cluster energy and
$p$ is the track momentum.  The $p_T$ measurement
of an electron is based on calorimeter energy information
and track position.

Muons are reconstructed within $|\eta| < 2.0$,
must be spatially matched to a track from the central 
tracking system, and are required to have matched sets of wire and scintillator hits 
before and after the muon toroid.
The detector support structure limits the muon system coverage in the region
$|\eta| < 1.1$ and $4.25 < \phi < 5.15$~\cite{d0coord};
in this region a single set
of matched wire and scintillator hits is required.  
Additionally, muons must be isolated such that the $p_T$ sum of other tracks
in a cone $\mathcal{R}=\sqrt{(\Delta\eta)^2 + (\Delta\phi)^2} < 0.5$
is $< 2.5$~GeV and calorimeter energy within $0.1 < \mathcal{R} < 0.4$ is $< 2.5$~GeV.

The \MET\ is determined based on the calorimeter energy deposition
distribution with respect to the interaction vertex.
It is corrected for the electromagnetic or jet energy scale, as appropriate,
and the $p_T$ of muons.

Signal acceptances and background processes are studied with a detailed 
Monte Carlo (MC) simulation based on {\sc pythia}~\cite{pythia} in conjunction with
the {\sc cteq6l1}~\cite{cteq} parton distribution functions, with detector simulation carried out 
by {\sc geant}~\cite{geant}.
The $Z$ boson $p_T$ spectrum in $Z/\gamma^*\to\ell\ell$ MC events is adjusted
to match data~\cite{zpt}.

For each final state, we require the highest $p_T$ (leading) lepton to 
have $p_T > 25$~GeV,
the trailing lepton to have $p_T > 15$~GeV, and the leptons to be of opposite 
charge. Both charged leptons are required to originate from the same vertex. 
The leptons must also have a minimum separation in $\eta$-$\phi$ space 
of $\mathcal{R}_{ee} > 0.8$ in the $ee$ channel or 
$\mathcal{R}_{e\mu/\mu\mu} > 0.5$ in the $e\mu$ and $\mu\mu$ channels,
in order to prevent overlap of the lepton isolation cones.

\begin{figure}
\begin{center}
\mbox{
\includegraphics[width=1.62in]{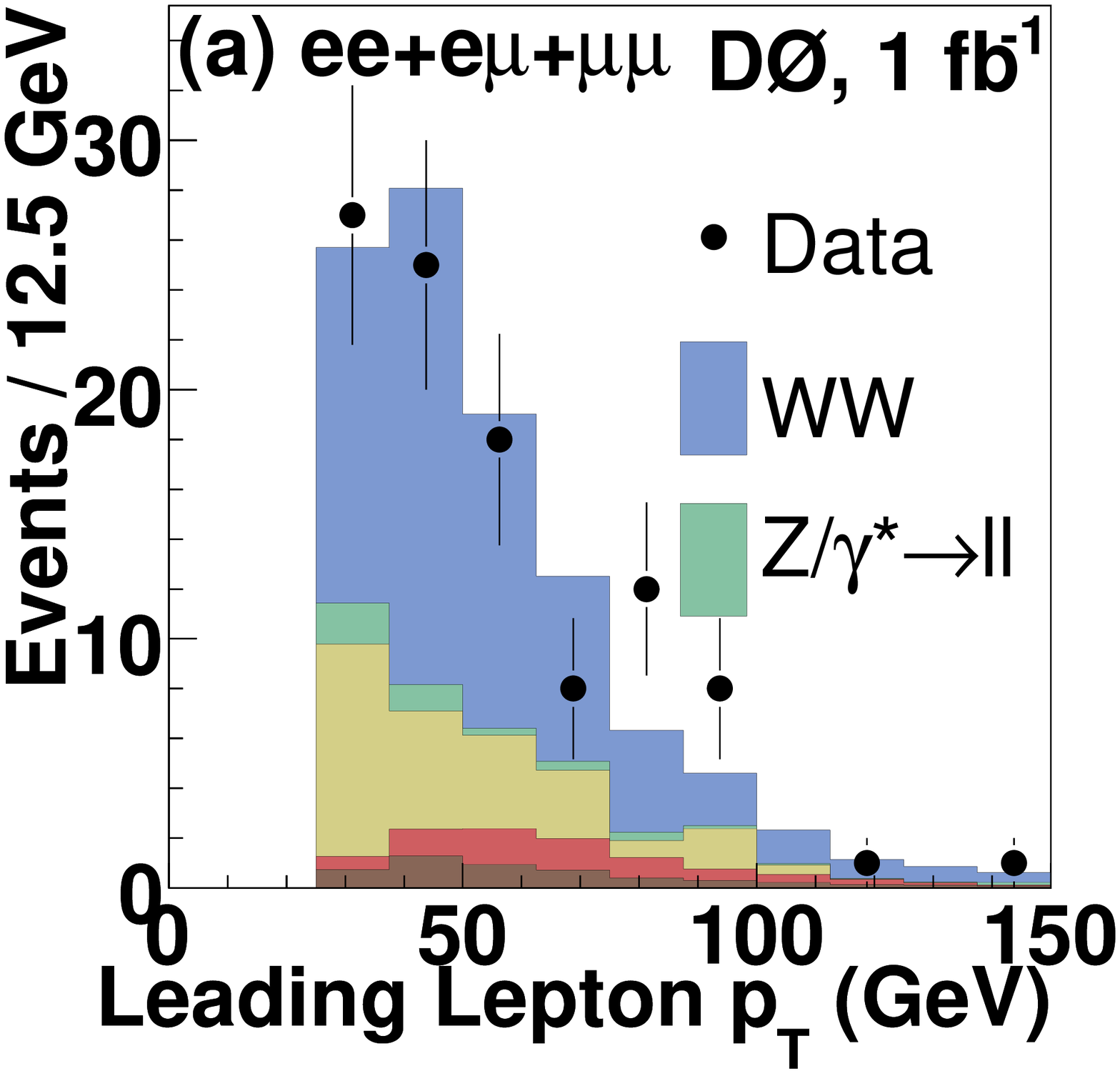} \hfill
\includegraphics[width=1.62in]{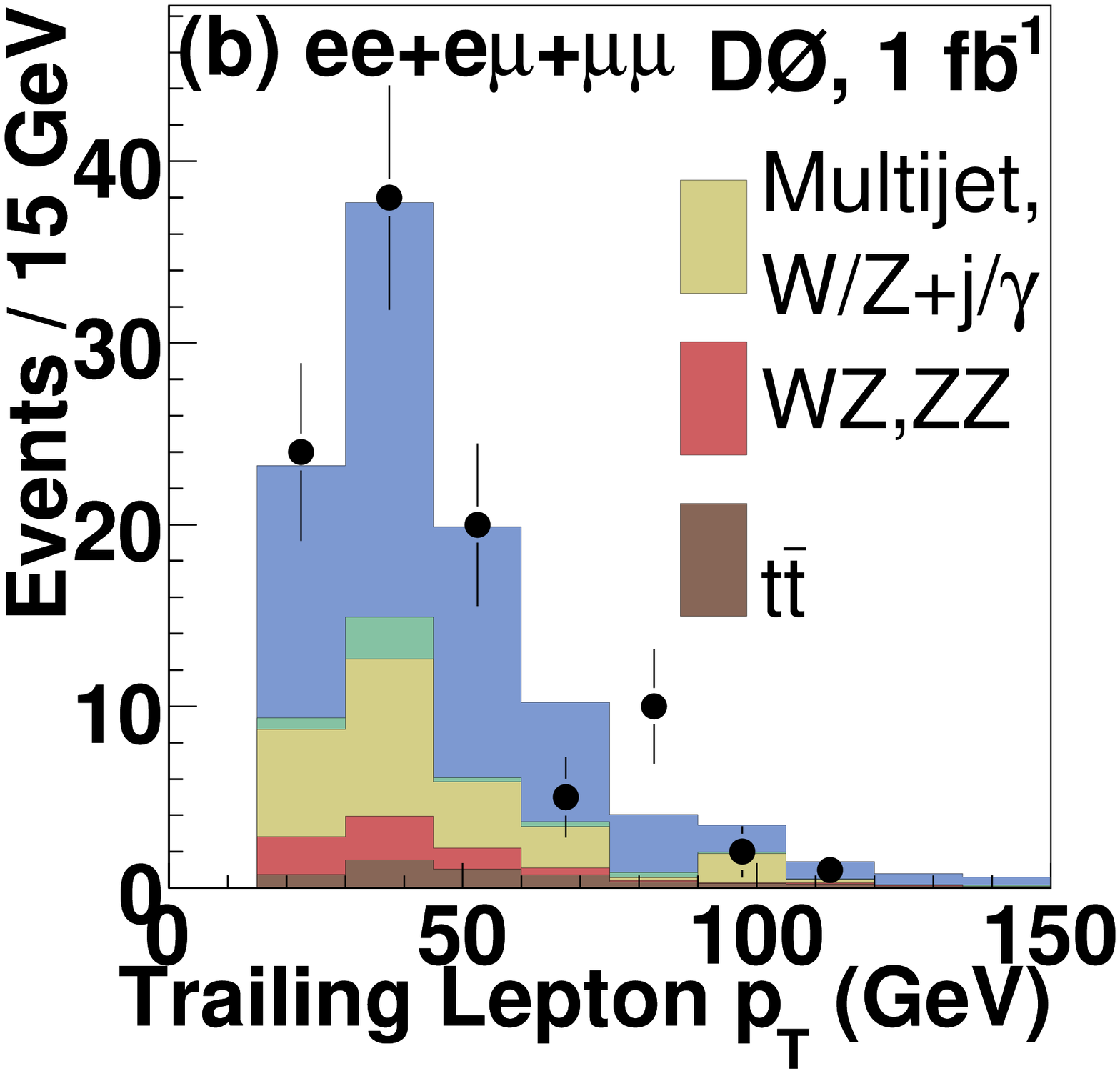}
}
\caption{\label{fig:leptonpt}
Distributions of the (a) leading and (b) trailing lepton $p_T$
after final selection, combined for all channels ($ee+e\mu+\mu\mu$).
Data are compared to estimated signal, $\sigma(WW)=12$~pb, and background sum.
}
\end{center}
\end{figure}

Background contributions to $WW$ production from $W$+jets 
and multijet production are estimated from the data.
Those from $Z/\gamma^*\to\ell\ell$, $t\bar{t}$, $WZ$, $W\gamma$, and $ZZ$
are estimated from the MC simulation. 

After the initial event selection, the dominant background in each channel is $Z/\gamma^*\to\ell\ell$
($\ell=e,\mu,\tau$).  Much of this background is removed by requiring 
$\slash\hspace{-0.65em}E_T > 45$ ($ee$), 20 ($e\mu$), or 35 ($\mu\mu$)~GeV.
For the $ee$ channel, we require 
$\slash\hspace{-0.65em}E_T > 50$~GeV if $|M_Z - m_{ee}| < 6$~GeV 
to further reduce the $Z/\gamma^*\to\ell\ell$ background. 
In events containing muons, a requirement on the azimuthal separation ($\Delta\phi$)
between the leptons is more effective at reducing 
the $Z/\gamma^*\to\ell\ell$ background than an invariant mass requirement, 
since the momentum resolution for high $p_T$ muons is poorer than the 
calorimeter energy resolution for electrons.
The $e\mu$ channel additionally requires \MET\ $> 40$ (instead of 20)~GeV if $\Delta\phi_{e\mu} > 2.8$,
and the $\mu\mu$ channel requires $\Delta\phi_{\mu\mu} < 2.45$.

Mismeasurement of the muon momentum can lead to spurious \MET\ which is collinear with the muon 
direction.  Especially in the $\mu\mu$ channel, mismeasurement of the muon momentum can
allow $Z$ boson events to satisfy the \MET\ requirement. 
To suppress these events in the $\mu\mu$ channel, we require that
the track for each muon candidate include at least one silicon microstrip tracker 
hit, for better momentum resolution, and that the azimuthal 
angle between each muon and the direction of the \MET\ satisfies 
\mbox{$|\cos(\Delta\phi_{\slash\hspace{-0.45em}E_T,\mu})| < 0.98$}.
 
A second background is $t \bar{t}$ production followed by the leptonic decay of $W$ bosons. 
This background can be suppressed by requiring
$q_T = | \overrightarrow{p_{T}}_{\ell} + \overrightarrow{p_{T}}_{\ell^\prime} + 
\overrightarrow{\slash\hspace{-0.65em}E_T} | < 20$ ($ee$), 25 ($e\mu$), or 16 ($\mu\mu$)~GeV. 
This quantity is the $p_T$ of the $WW$ system and is expected to be small 
for signal events.
However, for $t\bar{t}$ production and other background processes, $q_T$ 
can be large, so this variable is a powerful discriminant against 
these backgrounds.

The $W\gamma$ process is a background for only the $ee$ and $e\mu$ channels, since 
the probability for a photon to be misidentified as a muon is negligible.
We determine the probability that a photon is misidentified as an electron with photons from 
$Z/\gamma^*\to ee\gamma$ decays
and use it to correct the MC-based prediction of the $W\gamma$ background.
The $W$+jets background, in which a jet is misidentified as an 
electron or muon, is determined 
from the data by selecting dilepton samples with loose and tight lepton requirements and
setting up a system of linear equations to solve for the $W$+jet 
backgrounds after all event selection cuts, similar to the multijet 
background estimation performed in~\cite{mmethod}.
The multijet background contains jets that are misidentified as the two 
lepton candidates. It is represented by a data sample where the 
reconstructed leptons fail the lepton quality requirements. This sample 
is normalized with a factor determined at preselection using 
like-charged lepton events.
It is assumed that misidentified jets result in randomly assigned charge signs.

The leptonic decay of $WZ$ and $ZZ$ events can mimic the $WW$ signal when one or more
of the charged leptons is not reconstructed and instead contributes to \MET.  
The $ZZ\to\ell\ell\nu\nu$ process is suppressed 
by the $|M_Z-m_{ee}|$ or $\Delta\phi_{\ell\ell^\prime}$ cut.

\begin{table}
\caption{\label{tab:cutflow}
Numbers of signal and background events expected and number of events observed
after the final event selection in each channel.  Negligible contributions are not shown.
Uncertainties include 
contributions from statistics and lepton selection efficiencies.  
}
\begin{tabular}{lr@{$\,\pm\,$}lr@{$\,\pm\,$}lr@{$\,\pm\,$}l}
\hline\hline
Process & \multicolumn{2}{c}{$ee$} & \multicolumn{2}{c}{$e\mu$} & \multicolumn{2}{c}{$\mu\mu$} \\
\hline
$Z/\gamma^*\to ee/\mu\mu$	& 0.27 & 0.20 	& 2.52 & 0.56 	& 0.76 & 0.36 \\
$Z/\gamma^*\to\tau\tau$		& 0.26 & 0.05 	& 3.67 & 0.46 	& \multicolumn{2}{c}{---} \\
$t\bar{t}$			& 1.10 & 0.10	& 3.79 & 0.17	& 0.22 & 0.04 \\
$WZ$				& 1.42 & 0.14 	& 1.29 & 0.14 	& 0.97 & 0.11 \\
$ZZ$				& 1.70 & 0.04	& 0.09 & 0.01 	& 0.84 & 0.03 \\
$W\gamma$			& 0.23 & 0.16 	& 5.21 & 2.97 	& \multicolumn{2}{c}{---} \\
$W+\text{jet}$			& 6.09 & 1.72 	& 7.50 & 1.83	& 0.12 & 0.24 \\
Multijet			& 0.01 & 0.01	& 0.14 & 0.13 	& \multicolumn{2}{c}{---} \\
\hline
$WW\to\ell\ell^\prime$		&10.98 & 0.59 	&39.25 & 0.81  	& 7.18 & 0.34 \\
$WW\to\ell\tau/\tau\tau\to\ell\ell^\prime$& 1.40 & 0.20  & 5.18 & 0.29 	& 0.71 & 0.10 \\
\hline
Total expected			&23.46 & 1.90 	&68.64 & 3.88 	&10.79 & 0.58 \\ 
Data		& \multicolumn{2}{c}{22} & \multicolumn{2}{c}{64} & \multicolumn{2}{c}{14} \\
\hline\hline
\end{tabular}
\end{table}

For each channel, the exact selection requirements
on \MET, $q_T$, and $|M_Z-m_{ee}|$ or $\Delta\phi_{\ell\ell^\prime}$ 
are chosen by performing a grid search on signal MC and expected background, 
minimizing the combined statistical and systematic uncertainty on the expected
cross section measurement.  The final lepton 
$p_T$ distributions are shown in Fig.~\ref{fig:leptonpt}~\cite{epaps}.

The overall detection efficiency for signal events is determined using MC with full
detector, trigger, and reconstruction simulation 
and is 7.18\% ($ee$), 13.43\% ($e\mu$), and 5.34\% ($\mu\mu$) for
$WW\to\ell\nu\ell^\prime\nu$ ($\ell,\ell^\prime=e,\mu$) decays and
2.24\% ($ee$), 4.36\% ($e\mu$), and 1.30\% ($\mu\mu$) for
$WW\to\tau\nu\ell\nu/\tau\nu\tau\nu\to\ell\ell^\prime+n\nu$ decays.
The numbers of estimated signal and background events and the number of observed events
for each channel after the final event selection are summarized in Table~\ref{tab:cutflow}.
The observed events are statistically consistent with the SM expectation in each channel.
Assuming the $W$ boson and $\tau$ branching ratios from~\cite{pdg},
the observations in data correspond to 
$\sigma(p\bar{p}\to WW) = 10.6 \pm 4.6 \text{ (stat)} \pm 1.9 \text{ (syst)} \pm 0.7 \text{ (lumi)}$ pb
in the $ee$ channel, 
$10.8 \pm 2.2 \pm 1.1 \pm 0.7$ pb
in the $e\mu$ channel, and
$16.9 \pm 5.7 \pm 1.4 \pm 1.0$ pb
in the $\mu\mu$ channel.
The dominant sources of systematic uncertainty for each channel are
the statistics associated with the estimation of the $W$+jet contribution in the $ee$ channel, the
photon misidentification probability used to estimate the $W\gamma$ contribution in the $e\mu$ channel,
and the MC statistics for backgrounds in the $\mu\mu$ channel~\cite{epaps}.  

The cross section measurements in the individual channels are combined
using the best linear unbiased estimator (BLUE) method~\cite{BLUE} yielding:
$\sigma(p\bar{p}\to WW) = 11.5\ \pm  2.1\text{ (stat + syst)}\ \pm 0.7 \text{ (lumi) pb}$.
The standard model calculation of the $WW$ production
cross section at the Tevatron center of mass energy 
is $12.0\pm0.7$~pb~\cite{theory}.

The TGCs that govern $WW$ production can be parameterized by a general 
Lorentz-invariant Lagrangian with fourteen independent complex coupling 
parameters, seven each for the $WW\gamma$ and $WWZ$ vertices~\cite{Hagiwara}.
Limits on the anomalous couplings are often obtained by taking the parameters to be
real, enforcing electromagnetic gauge invariance, and assuming 
charge conjugation and parity
invariance, reducing the number of independent couplings to five:
$g^Z_1$, $\kappa_Z$, $\kappa_\gamma$, $\lambda_Z$, and $\lambda_\gamma$
(using notation from~\cite{Hagiwara}).
In the SM, $g^Z_1 = \kappa_Z = \kappa_\gamma = 1$ and $\lambda_Z =
\lambda_\gamma = 0$. 
The couplings that are non-zero in the SM  are often
expressed in terms of their deviation from the SM values, e.g.
$\Delta g^Z_1 \equiv g^Z_1 - 1$. 
Enforcing $SU(2)_L\otimes U(1)_Y$ symmetry introduces
two relationships between the remaining parameters: $\kappa_Z = g_1^Z - 
(\kappa_\gamma - 1) \mathrm{tan}^2 \theta_W$ and $\lambda_Z = 
\lambda_\gamma$,  reducing the number of
free parameters to three~\cite{constraint}.
Alternatively, enforcing equality between the $WW\gamma$ and $WWZ$
vertices ($WW\gamma$=$WWZ$) such that $\kappa_\gamma=\kappa_Z$, $\lambda_\gamma=\lambda_Z$,
and $g^Z_1=1$ reduces the number of free parameters to two.

One effect of introducing anomalous coupling parameters into the
SM Lagrangian is an enhancement of the cross section for the
$q\bar{q} \to Z/\gamma^* \to W^+W^-$ process, which 
leads to unphysically large cross sections at high energy.
Therefore, the anomalous couplings must vanish as 
the partonic center of mass energy $\sqrt{\hat{s}} \to \infty$. 
This is achieved by introducing a dipole form factor for an 
arbitrary coupling  $\alpha$ ($g^Z_1$, $\kappa_V$, or $\lambda_V$):
$\alpha(\hat{s}) = \alpha_0/(1 + \hat{s}/\Lambda^2)^2$,
where the form factor scale $\Lambda$ is set by new physics, and limits
are set in terms of $\alpha_0$. 
Unitarity constraints provide an upper limit for each coupling that is
dependent on the choice of $\Lambda$.
For this analysis we use 
$\Lambda=2$ TeV, the approximate center of mass energy of the Tevatron. 
 
The leading order MC event generator by Hagiwara, Woodside, and Zeppenfeld~\cite{Hagiwara}
is used to predict the changes in $WW$ production cross section and kinematics
as coupling parameters are varied about their SM values.
At each point on a grid in TGC parameter space,
events are generated and passed through 
a parameterized simulation of the D0 detector that is tuned 
to data.  To enhance the sensitivity to anomalous couplings, 
events are sorted by lepton $p_T$ into a two-dimensional histogram,
using leading and trailing lepton
$p_T$ values in the $ee$ and $\mu\mu$ channels, and 
$e$ and $\mu$ $p_T$ values in the $e\mu$ channel.
For each bin in lepton $p_T$ space, the expected number of $WW$ events produced
is parameterized by a quadratic function in three-dimensional 
\mbox{($\Delta\kappa_\gamma$,$\lambda_\gamma$,$\Delta g_1^Z$)} space
or two-dimensional \mbox{($\Delta\kappa$,$\lambda$)} space, as
appropriate for the TGC relationship scenario under study.
In the three-dimensional case, coupling parameters are investigated in pairs, with the third
parameter fixed to the SM value.
A likelihood surface is generated by considering all channels simultaneously,
integrating over the signal, background, and luminosity
uncertainties with Gaussian distributions using the same methodology as
that used in previous studies~\cite{d0wwac}.

\begin{figure} [t]
\begin{center}
\mbox{
\includegraphics[width=1.62in]{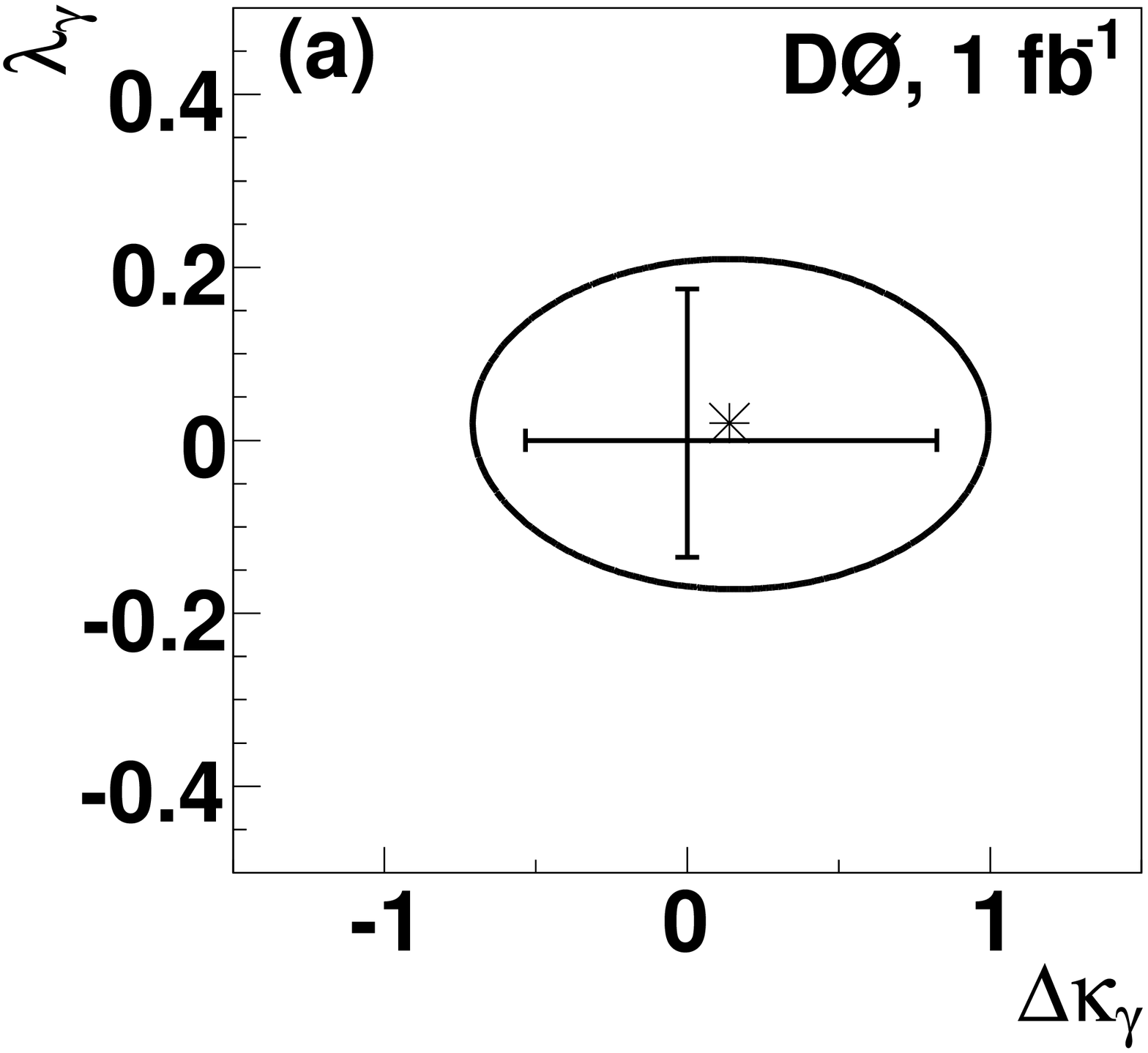} \hfill
\includegraphics[width=1.62in]{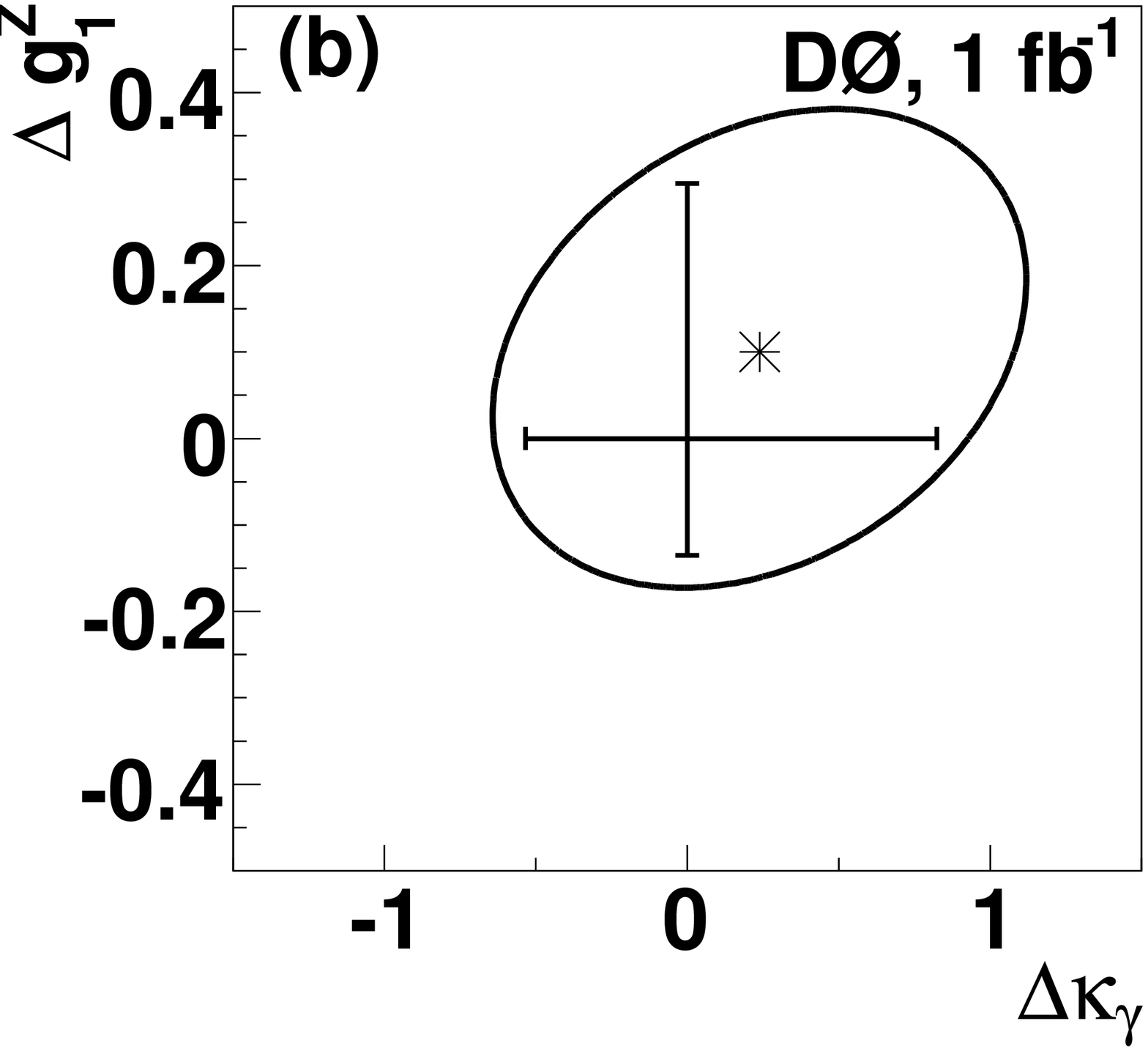}
}
\mbox{
\includegraphics[width=1.62in]{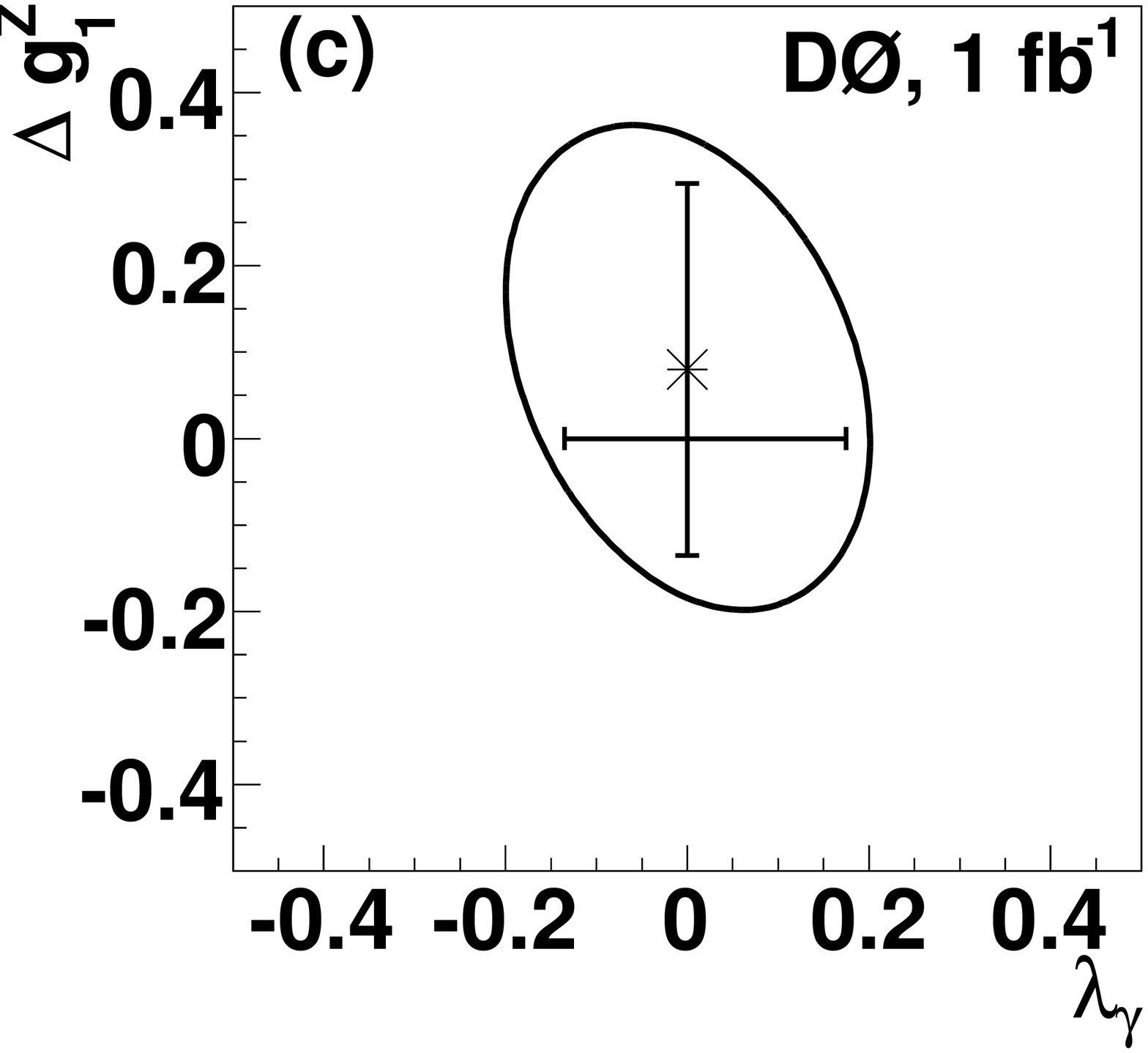} \hfill
\includegraphics[width=1.62in]{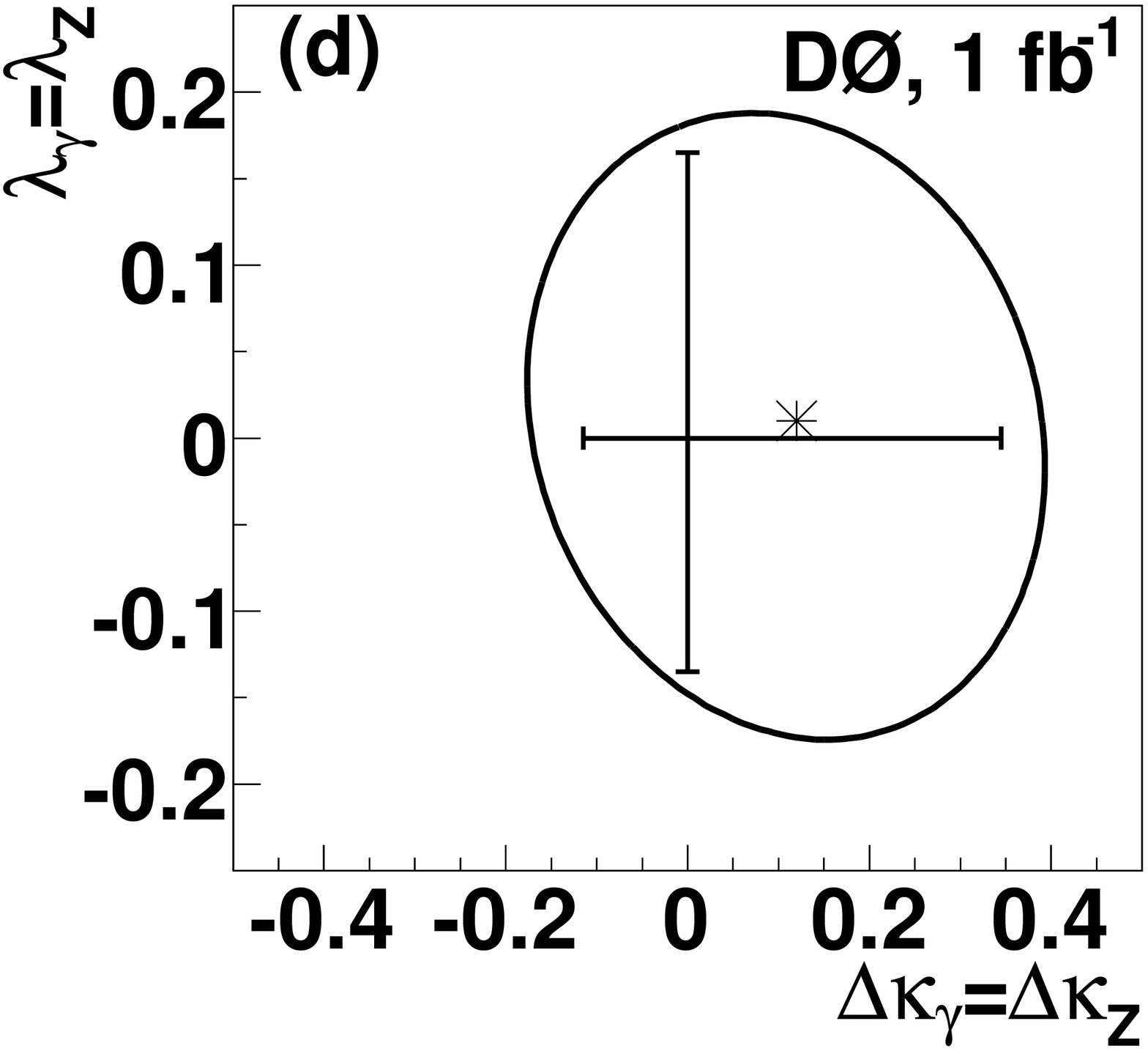}
}
\caption{\label{fig:aclimits}
One and two-dimensional 95\%~C.L.\ limits
when enforcing $SU(2)_L \otimes U(1)_Y$ symmetry at $\Lambda = 2$ TeV, 
for
(a) $\Delta\kappa_\gamma$ vs. $\lambda_\gamma$,
(b) $\Delta\kappa_\gamma$ vs. $\Delta g_1^Z$, and
(c) $\lambda_\gamma$ vs. $\Delta g_1^Z$,
each when the third free coupling is set to its
SM value;
limits when enforcing the $WW\gamma$=$WWZ$ constraints are shown in (d).
The curve represents the two-dimensional 95\%~C.L.\ contour
and the ticks along the axes represent the one-dimensional
95\%~C.L.\ limits.
An asterisk ($+ \hspace{-0.8em} \times$) marks the point 
with the highest likelihood in the two-dimensional plane.
}
\end{center}
\end{figure}

The one-dimensional 95\%~C.L.\ limits for $\Lambda = 2$~TeV are determined to be 
$-0.54 < \Delta\kappa_\gamma < 0.83$,
$-0.14 < \lambda_\gamma=\lambda_Z < 0.18$, and
$-0.14 < \Delta g_1^Z < 0.30$
under the $SU(2)_L\otimes U(1)_Y$-conserving constraints,
and $-0.12 < \Delta\kappa_\gamma=\Delta\kappa_Z < 0.35$,
with the same $\lambda$ limits as above,
under the $WW\gamma$=$WWZ$ constraints.
One- and two-dimensional 95\%~C.L.\ limits are shown in Fig.~\ref{fig:aclimits}.

In summary, we have made the most precise measurement of
$WW$ production at a hadronic collider to date,
$\sigma(p\bar{p}\to WW) = 11.5\ \pm  2.1\text{ (stat + syst)}\ \pm 0.7 \text{ (lumi) pb}$,
using 1 fb$^{-1}$ of data at the D0 experiment.
This result is consistent with the SM prediction and
previous Tevatron results~\cite{d0RunII,theory,cdfwwxsec}.  
The selected event kinematics are used to significantly
improve previous limits on anomalous TGCs from $WW$ production at the Tevatron,
reducing the allowed 95\% C.L. interval for $\lambda_\gamma=\lambda_Z$ and
$\Delta\kappa_\gamma=\Delta\kappa_Z$ by nearly a factor of two~\cite{d0wwac,cdfwwac}.

\begin{acknowledgments}
%
We thank the staffs at Fermilab and collaborating institutions, 
and acknowledge support from the 
DOE and NSF (USA);
CEA and CNRS/IN2P3 (France);
FASI, Rosatom and RFBR (Russia);
CNPq, FAPERJ, FAPESP and FUNDUNESP (Brazil);
DAE and DST (India);
Colciencias (Colombia);
CONACyT (Mexico);
KRF and KOSEF (Korea);
CONICET and UBACyT (Argentina);
FOM (The Netherlands);
STFC and the Royal Society (United Kingdom);
MSMT and GACR (Czech Republic);
CRC Program, CFI, NSERC and WestGrid Project (Canada);
BMBF and DFG (Germany);
SFI (Ireland);
The Swedish Research Council (Sweden);
CAS and CNSF (China);
and the
Alexander von Humboldt Foundation (Germany).

\end{acknowledgments}

\vfill


\clearpage

\section{Supplemental Material}

\begin{figure}[!b]
\begin{center}
\mbox{
\includegraphics[width=1.62in]{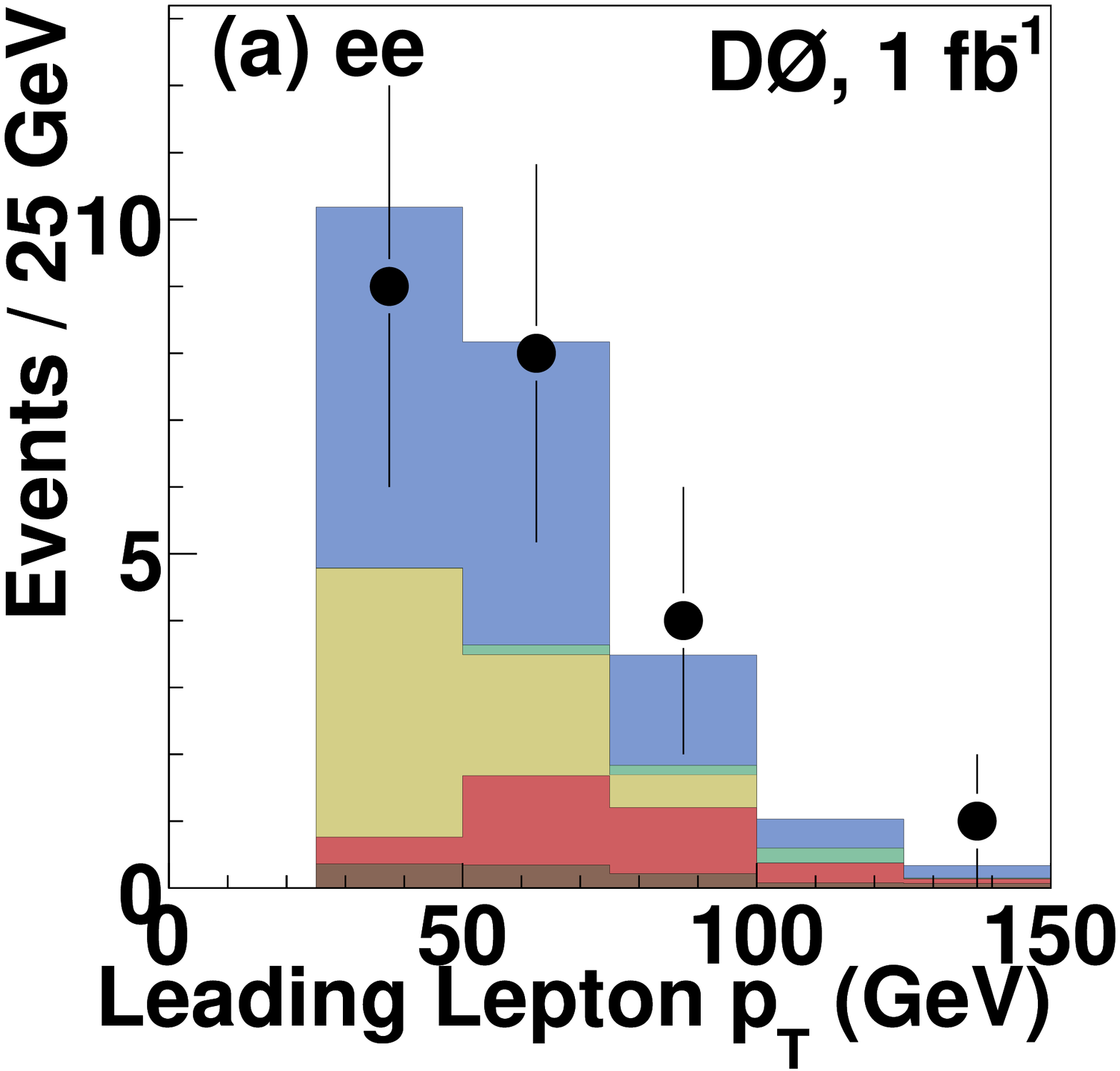} \hfill
\includegraphics[width=1.62in]{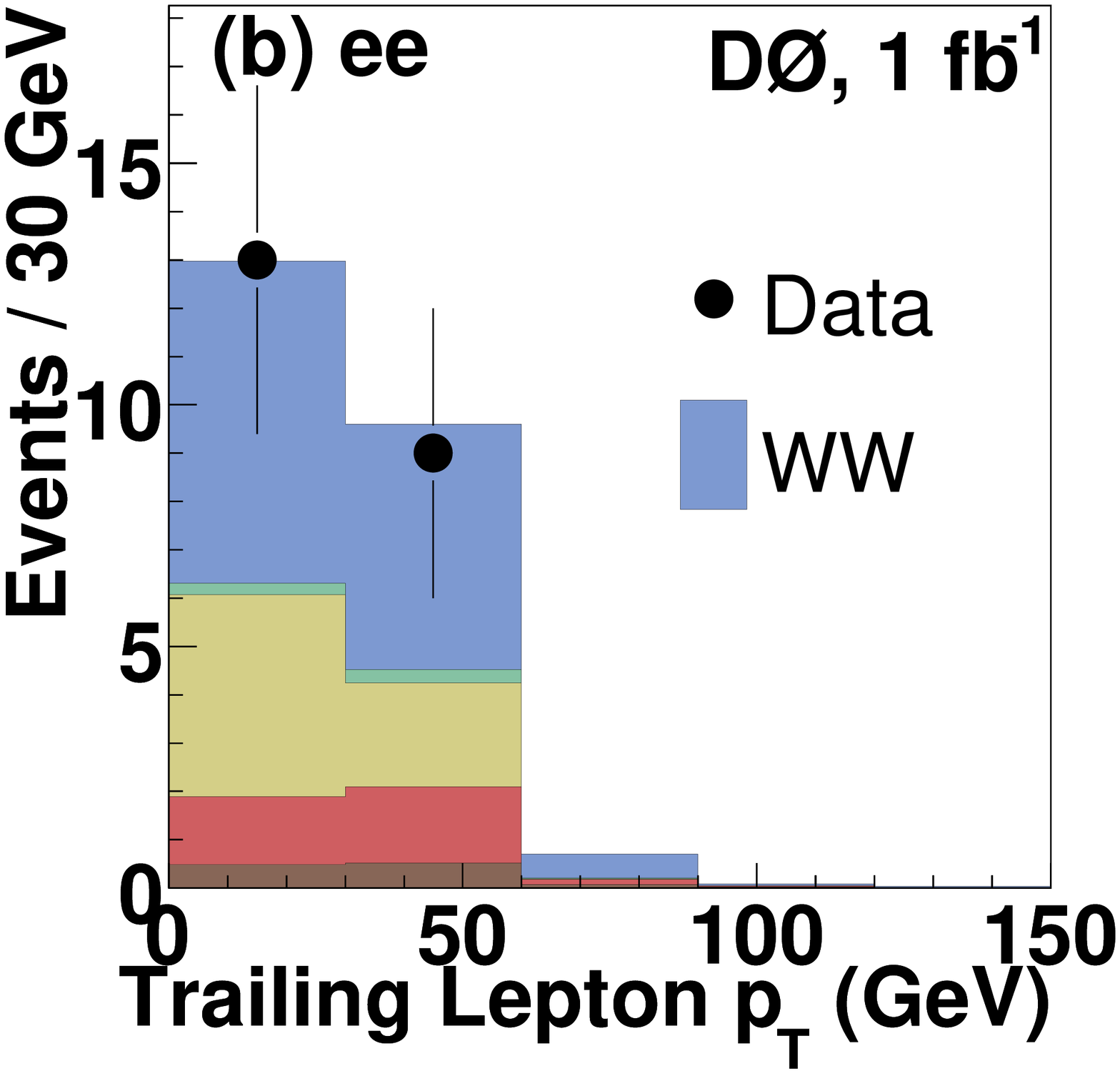}
}
\mbox{
\includegraphics[width=1.62in]{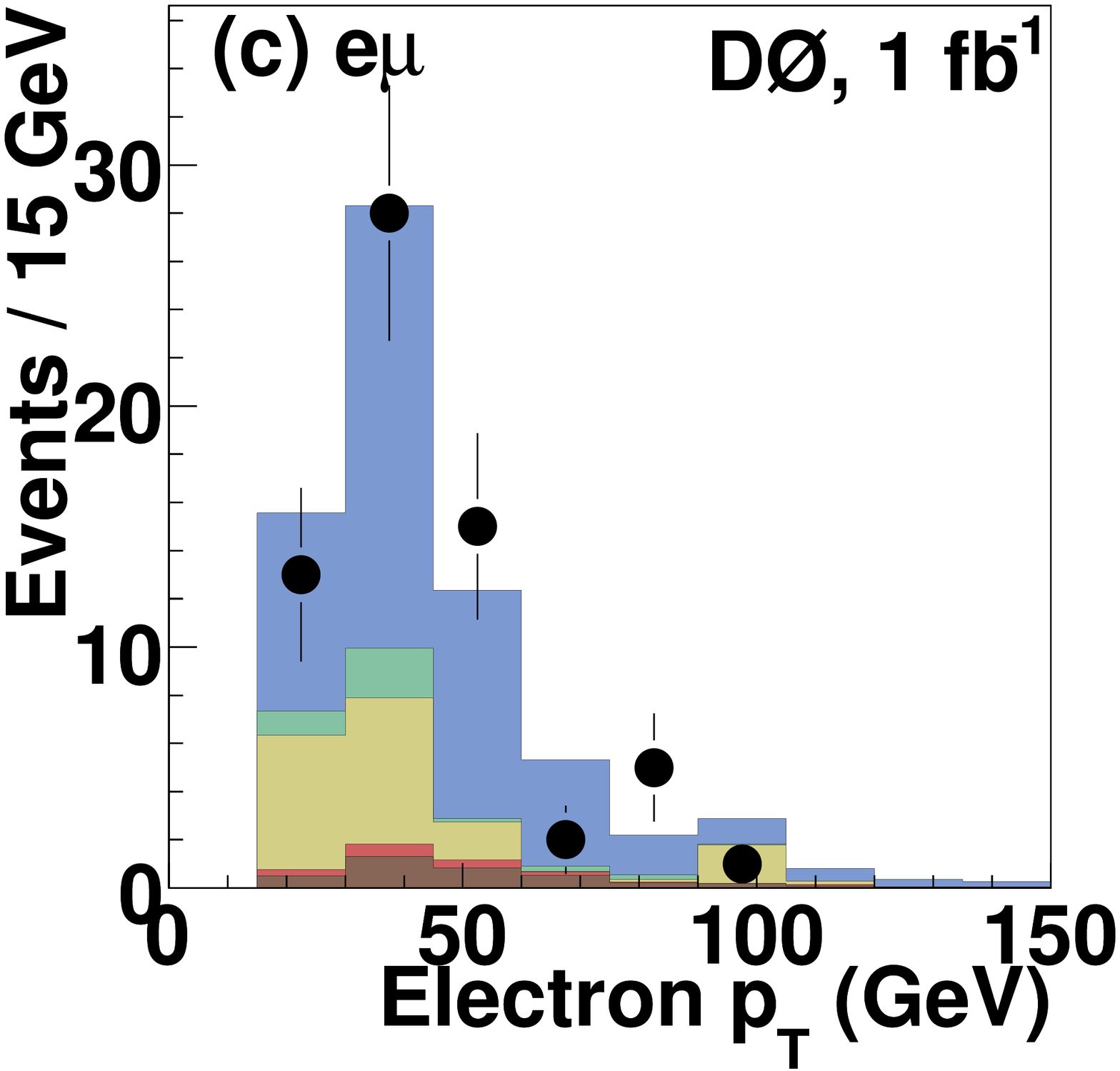} \hfill
\includegraphics[width=1.62in]{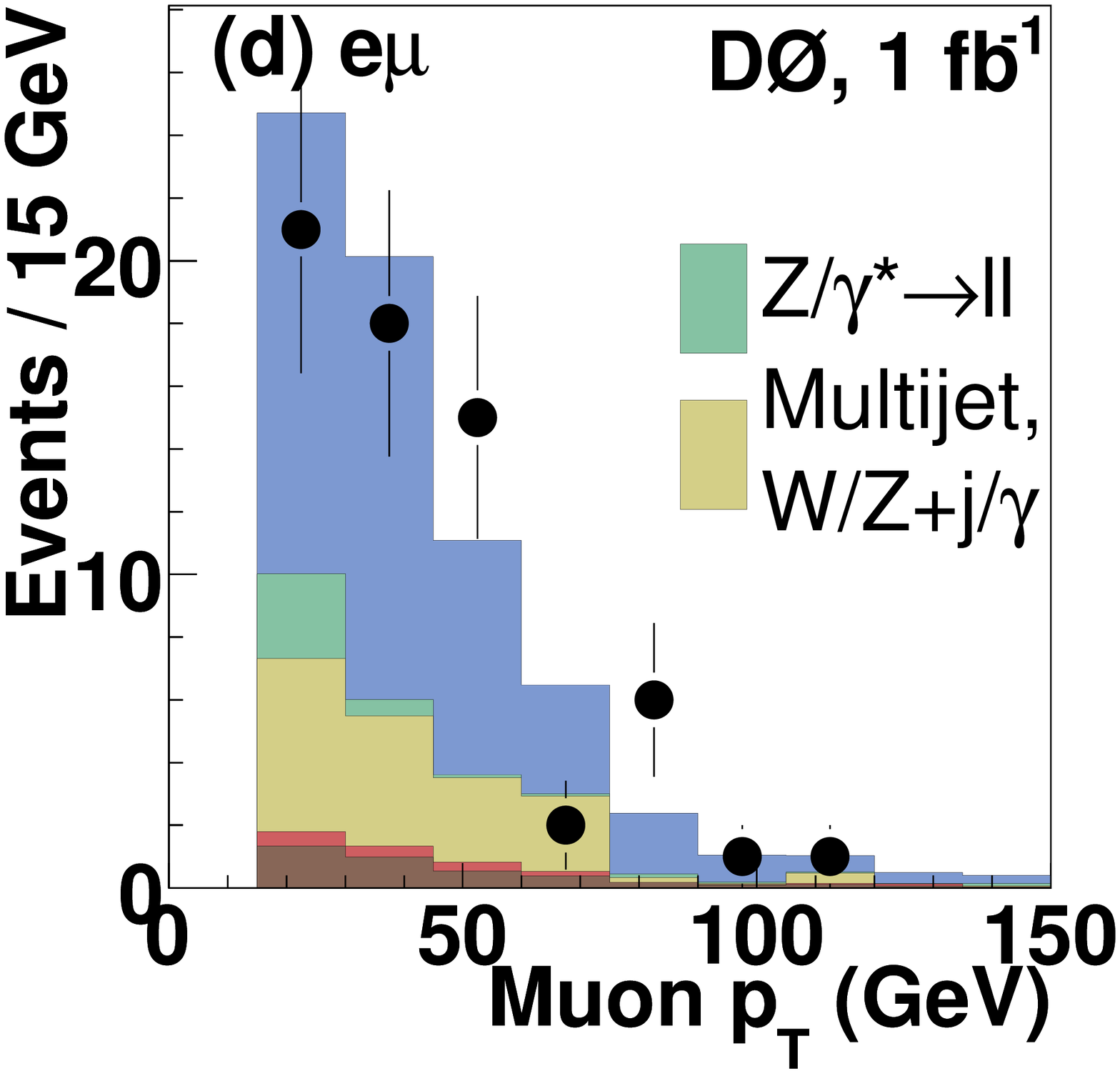}
}
\mbox{
\includegraphics[width=1.62in]{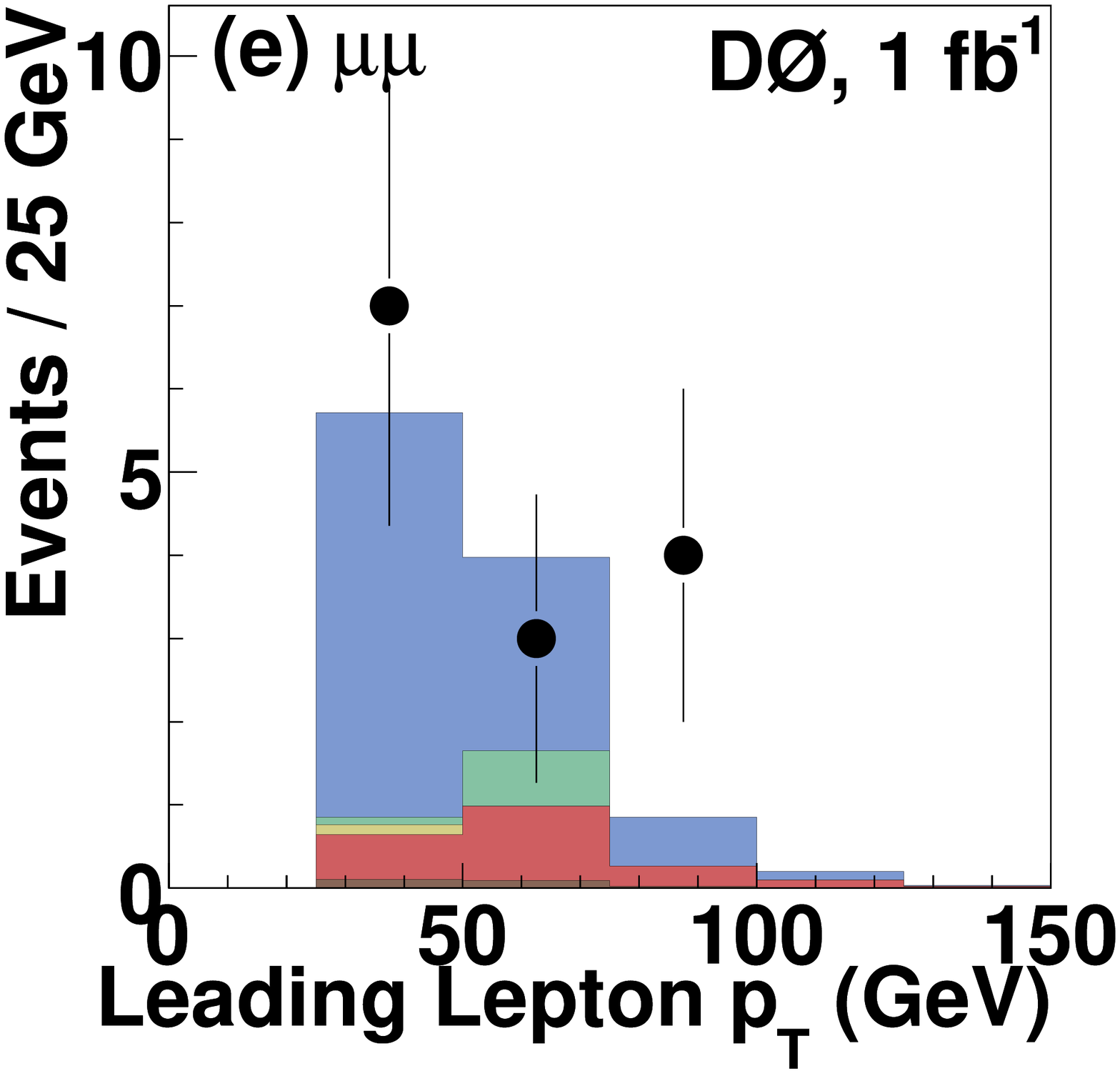} \hfill
\includegraphics[width=1.62in]{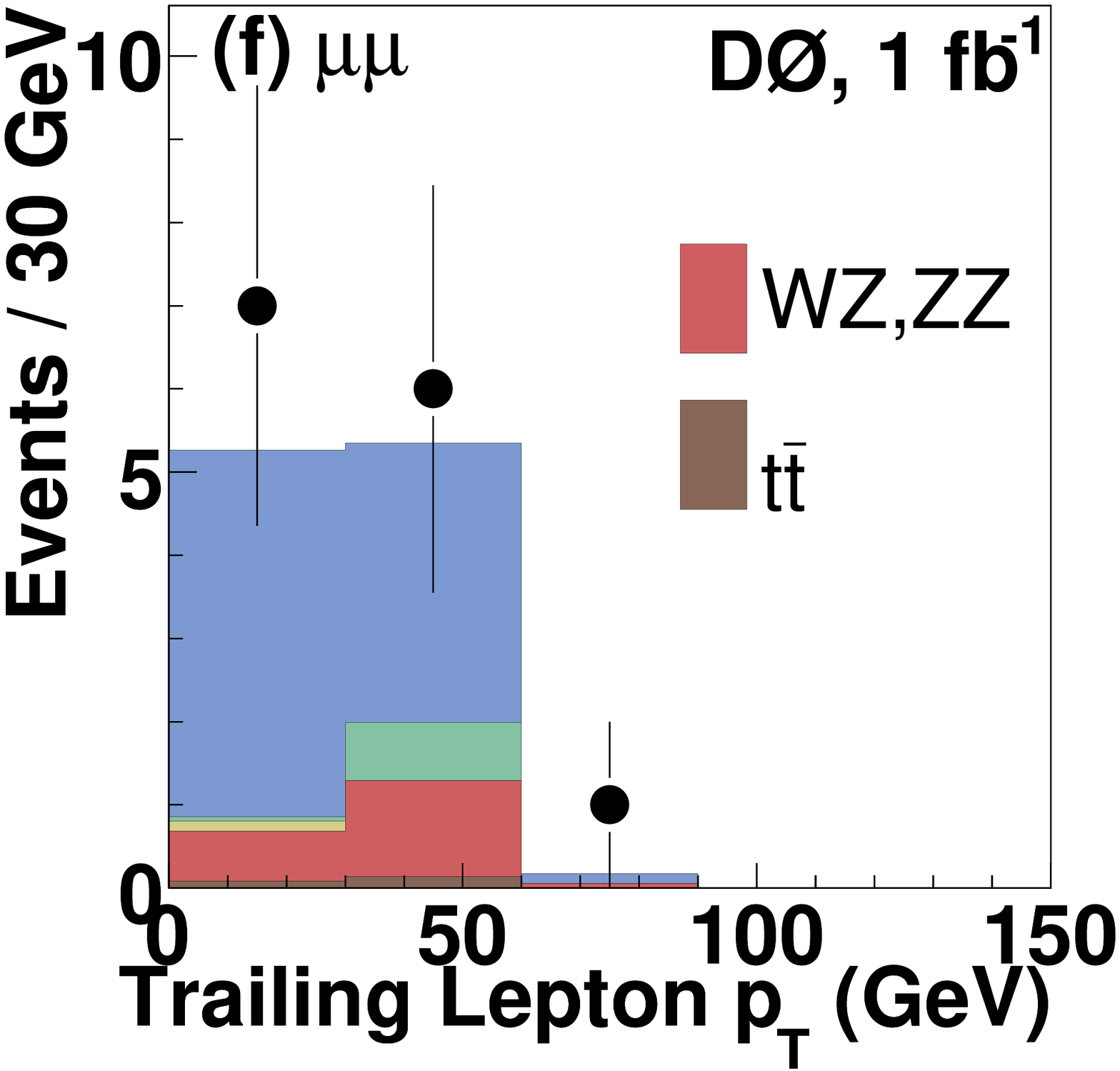}
}
\caption{\label{fig:metqt}
Distributions of the (a) leading and (b) trailing electron $p_T$ in the $ee$ channel,
(c) electron and (d) muon $p_T$ in the $e\mu$ channel, and 
(e) leading and (f) trailing muon $p_T$ in the $\mu\mu$ channel after final selection.
Data are compared to estimated signal, $\sigma(WW)=12$~pb, and background sum.
}
\end{center}
\end{figure}

The final lepton $p_T$ distributions for individual analysis channels
are shown in Figure~\ref{fig:metqt}.
The uncertainty from each systematic source as a percentage of the final
background estimate for each channel is provided in Table~\ref{tab:syst}.

\begin{table}
\caption{\label{tab:syst}
Uncertainty from each systematic source as a percentage of the final background
estimate for each channel.
}
\begin{tabular}{p{2.0in}rrr}
\hline\hline
Systematic Source 			& $ee$   & $e\mu$ & $\mu\mu$ \\
\hline
MC Statistics 				& 2.6\%  &  3.6\% & 12.8\% \\
MC Cross Section			& 2.6\%  &  2.2\% &  2.5\% \\
MC Corrections				& 2.0\%  &  6.1\% &  2.0\% \\
Electron ID				& 0.3\%  &  0.3\% &  ---   \\
Muon ID					& ---    &  0.4\% &  3.0\% \\
$\gamma\to e$ Mis-ID Rate		& 1.2\%  & 12.1\% &  ---   \\
$W+$jet Statistics			& 15.5\% &  7.5\% &  6.8\% \\
$W+$jet $e$ Efficiency \& Mis-ID Rate	& 3.4\%  &  2.9\% &  ---   \\
$W+$jet $\mu$ Efficiency \& Mis-ID Rate	& ---    &  1.4\% &  2.1\% \\
\hline
Total Background Uncertainty		& 16.4\% & 16.4\% & 15.3\% \\
\hline\hline
\end{tabular}
\end{table}

In Table~\ref{tab:syst}, the ``MC Statistics'' uncertainty is based on the number of MC events used
to estimate the number of background events after final selection.  The ``MC Cross
Section'' uncertainty is based on the PDF and scale uncertainties
associated with the theoretical cross sections used to scale the
MC-driven background estimations.  The ``MC Corrections'' uncertainty
accounts for data-driven reweighting of the MC that corrects for general
event characteristics such as the primary vertex $z$-position and
instantaneous luminosity distributions.  
The ``$\gamma\to e$ Mis-ID Rate'' uncertainty is driven
by statistics in the data sample used to determine that misidentification rate.  
Sources of systematic uncertainty for the $W+$jet background estimation
are separated into categories based upon their correlations
across analysis channels.
The ``$W+$jet Statistics'' systematic is based on the number of
events in data that pass a series of lepton identification requirements
in each analysis channel, independently.
The ``$W+$jet Lepton Efficiency \& Mis-ID Rate''
uncertainties are based on measuring the rate at which a lepton passes
loose and tight identification in $Z/\gamma^*\to\ell\ell$ candidate events
in data and the rate at which a jet is misidentified as a lepton with loose or tight 
identification based on dijet data.

The statistical uncertainty due to the number of observed events in
each channel is the dominant source of final uncertainty.

\vfill

\end{document}